\nopagenumbers
\baselineskip=20pt

\input boxedeps.tex
\input boxedeps.cfg

\SetDefaultEPSFScale{600}

\centerline {\bf { Dirac Equation in the Background of the Nutku
Helicoid Metric}} \vskip .7truein \centerline {T. Birkandan,
M.Horta\c csu
 }
\vskip 1.1truein \centerline { Physics Department, Faculty of
Sciences and Letters} \vskip .1truein \centerline { ITU 34469
Maslak, Istanbul, Turkey } \vskip 1.3truein \centerline {\bf {
Abstract}} \vskip.3truein \noindent We study the solutions of the
Dirac equation in the background of the  Nutku helicoid metric. This
metric has curvature singularities, which necessitates imposing a
boundary to exclude this point.  We use the Atiyah-Patodi-Singer non
local spectral boundary conditions for both the four and the five
dimensional manifolds.

\vskip 1.2truein \noindent key words: APS boundary conditions,
solutions of Dirac equation \vskip .5truein \noindent {E-mail
addresses: hortacsu@itu.edu.tr, birkandant@itu.edu.tr}
\baselineskip=18pt \footline={\centerline {\folio}} \pageno=1
\vskip.4truein
 PACS number: 04.62.+v

 \vfill\eject
\noindent
 {\bf{Introduction}}
 \vskip.1truein
\noindent New interesting solutions for gravitational instantons
exist in the literature.  One of them is the helicoid solution of
Nutku [1,2 ].  Since this solution has curvature singularities, it
has not been studied extensively aside from three articles [3,4,5 ].
These three papers study the solutions of the Dirac equation using
this solution as a background. Our work on these solutions can be
extended by studying them in a manifold with a boundary. Taking a
domain that excludes the origin and infinity for the radial variable
cures the problems associated with  the infinities of these
solutions.

\noindent One of us (M.H.), with collaborators, had studied the
related problems in the presence of an instanton and a meron in the
past [6,7 ]. Similar problems have been studied recently for a
spherical bag [8 ]. In these three papers, to conserve self
adjointness, as well as chirality and charge conjugation, spectral
boundary conditions [9,10 ] were used. The spectral boundary
conditions were also extended to the Dirac equation with torsion [11
].

\noindent Here we study the solutions to the Dirac equation in the
background of the Nutku helicoid metric.  We use Euclidean signature
for our metric.  This makes our differential operator of the
elliptic type. We find that  although some of the solutions diverge
at the origin, they are normalizable  when integrated with the
measure $d\tau \sqrt{g} $ ,where $\tau$ is the  volume. We also see
that the solution diverges as  $x$ goes to infinity, since they are
expressed in terms of modified Mathieu functions. Forcing them to be
regular at the origin makes them divergent at infinity. Furthermore,
the metric used has curvature singularities at the origin. Although
one can define the index on certain noncompact manifolds, the
conditions of "bounded geometry" are needed  [12,13 ]. These are the
conditions we do not meet. Thus we have to restrict our space at a
boundary.  This necessitates specifying the boundary conditions.

\noindent We know that local boundary conditions can be used in even
dimensions, although, it was shown  [6 ] that the  spectral boundary
conditions  [9,10 ] are the only self adjoint one that also
conserves the $\gamma^5$ and the charge conjugation symmetry of the
Dirac operator. The chiral Dirac operator, however, requires
spectral boundary conditions in even dimensional spaces [14]. The
non-local, spectral boundary conditions should be used for the odd
dimensional case due to obstructions that were first pointed out by
Atiyah and Bott  [15 ]. Here we will study the problem both in odd
and even dimensions. We will use the spectral boundary conditions in
both cases.

\noindent The non-local spectral boundary conditions were introduced
by Atiyah-Patodi-Singer  while they were investigating the
Hirzebruch signature theorem for manifolds with boundaries
[15,16,17].  When a manifold has a boundary, we may use local
elliptic boundary conditions, like those of Dirichlet, Neumann or
Robin, if the manifold has even dimensions.  We may be sacrificing
the symmetries cited above, though.  If the manifold is odd
dimensional, however, we have to use the non-local, spectral
boundary conditions. Atiyah and Bott [15 ] found out that in general
there are topological obstructions to finding acceptable local
boundary conditions. As described in [15,16 ] the crucial point is
defining an acceptable elliptic boundary problem for the signature
operator whose index is the signature of the manifold. We call a
complex elliptic if we can define an elliptic operator on it. The
exterior algebra  can be split into two distinct elliptic complexes.
The first is the de Rham complex, which is related to the Euler
characteristic. The second is the signature complex [17 ]. Although
the de Rham complex admits local boundary conditions, it can be
shown that for the signature complex, as well as for the spin and
the Dolbeaux complexes, there does not exist a local boundary
condition so that this complex with the local boundary condition is
elliptic [18 ]. In the first order case, like the Dirac operator,
there is a natural pseudo-differential operator given by the
projection on the generalized eigenvectors with eigenvalues with
plus or minus real parts. This operator leads to a well posed
boundary value problem for the signature complex ( as well as the
spin and the Dolbeaux complexes.) This fact requires the use of the
spectral boundary conditions, which are non-local.

\noindent In defining the index of the operator, the presence of a
boundary necessitates extra terms in addition to the Euler number.
For the case without a boundary, knowing only the curvature suffices
to compute the index  [17 ]. One calculates the integral over the
whole manifold of the same characteristic classes.  If the manifold
has a boundary, then, we have to add  a term obtained from the
integration of  the Chern-Simons form, which is written in terms of
the connection, the curvature and the second fundamental form
determined from the normal to the boundary. Another term that should
be added is   the  $\eta$ invariant, which is calculated  from the
eigenvalues of the operator restricted to the boundary. We also add
a term proportional to the number of zero eigenvalue solutions of
this operator. The index is equal to the sum of these four terms.

\noindent Here we study the Dirac equation in the background of the
helicoid metric in five and four dimensions in  consecutive
sections. We do not attempt to calculate the index for our Dirac
operator. Our only aim is to make a sense out of this incomplete
metric by restricting the domain and using the appropriate boundary
conditions for this problem. \vskip .3truein

\vskip.3truein
 \noindent
 {\bf{Solutions of Dirac Equations}}
 \vskip.1truein
\noindent
 The relevance of the helicoid- catenoid metric, as a simplest minimal surface in Euclidean space resulting in a
 gravitational  instanton [1], for certain cosmological models is given in a recent preprint
 [21]. One can transform from the helicoid to the catenoid metric
 by a simple transformation [2].
 The importance of five dimensional spaces is also stressed in the same
 paper. The helicoid metric is also mentioned among the metrics
 related to {\bf{M}} theory by Gibbons et al [22]. This metric is
 also studied by Valent in [23,24,25].
 These are other reasons why we try to give a meaning to the solutions of Dirac equation in the background of the Nutku helicoid solution
 both in four and  five dimensions.  Here we briefly introduce the metric
  and write formally the Dirac equation in the background of this
  metric.

\noindent
   The Nutku helicoid metric is written for the four dimensional Euclidean
space. It is given as [2]

$$
{ ds}^{2}= {{1}\over{\sqrt{1+{{a^{2}}\over{r^{2}}}}}}%
[dr^{2}+(r^{2}+a^{2})d\theta ^{2}+\left( 1+{{a^{2}}\over{r^{2}}}sin
^{2}\theta \right) dy^{2}
-{{a^{2}}\over{r^{2}}}sin 2\theta dydz{+}\left( 1+{{a^{2}}\over{%
r^{2}}}cos ^{2}\theta \right)  dz^{2}] , \eqno{[1]}$$ where $0<r<
\infty$, $0 \leq \theta \leq 2\pi$, $y$ and $z$ are along the
Killing directions and will be taken to be periodic coordinates on a
2-torus. This is an example of a multi-center metric. This metric
reduces to the flat metric if we take $a=0$.

\noindent If we make the following transformation $ r=asinh x$, the
metric is written as
$$
ds_4^{2} ={{a^{2}}\over{2}}sinh 2x(dx^{2}+d\theta ^{2})  \ +\
{{2}\over{sinh 2x}}[(sinh ^{2}x+sin ^{2}\theta )dy^{2} -sin 2\theta
dydz+(sinh ^{2}x+cos ^{2}\theta )dz^{2}]. \eqno{[2]}$$ We use the NP
formalism  [26,27] in four Euclidean dimensions [28,29,30 ]. The
details of how the tetrad is chosen, how the $\gamma^{\mu} $
matrices are formed for this explicit case are given in [5 ].

\noindent We use the  Dirac operator $ i\gamma ^{\mu }\nabla _{\mu
}$ where
$$
\nabla _{\mu }=\partial _{\mu }-\Gamma _{\mu } . \eqno{[3]}$$
The
$\gamma$ matrices satisfy
$$
\{\gamma ^{\mu },\gamma ^{\nu }\}=2g^{\mu \nu }. \eqno{[4]}$$

\noindent The spin connection is written as $\Gamma _{\mu
}={{1}\over{4}}\gamma _{;\mu }^{\nu }\gamma _{\nu }.$

\noindent We can trivially extend this structure to five dimensions
The addition of the Euclidean "time component" to the previous
metric gives:

$$
ds^{2}=dt^{2}+ds_{4}^{2} . \eqno{[5]}$$ Going to five dimensions
does not increase the number of equations we have to solve, since in
both four and five dimensions we can use four component Dirac
spinors. We can not interpret the fifth dimension as time, since
this will bring problems with causality. If we use non-local
boundary conditions, these conditions are set for all times.
Abrikosov  has devised a new method  [31 ] which cures the causality
problem.  We could not, however, generalize this method to our case,
since our little Dirac equation still couples three components. We,
therefore, in both even and odd dimensional cases, study the
solution with the Euclidean metric. We include both cases to compare
and contrast the similarities and differences of these closely
related cases. \vskip.2truein \noindent
 {\bf{ Solutions in five dimensions}}
 \vskip.1truein
 \noindent
We start by studying the solutions to the Dirac equation in the
background of the Nutku helicoid metric in five dimensions. Our
motivation is to see whether one can find a domain where solutions
in the background of this singular metric can be defined.  We will
try to show that one can obtain a satisfactory solution to this
problem.

\noindent
  We write the system in the form $L \psi = \Lambda \psi $ and try to
obtain the solutions for the different components. Our aim is to
write the upper components in terms of the lower components. We
will impose the boundary conditions on the upper components in
terms of derivatives of the lower components below.

\noindent
 The equations read
$$
{{\sqrt{2}}\over{a\sqrt{sinh 2x}}}\{(\partial _{x}+i\partial
_{\theta })\Psi _{3}  \
+a[cos (\theta +ix)\partial _{y}+sin (\theta +ix)\partial _{z}]\Psi _{4}-%
{{{a\sqrt{sinh 2x}}\over{\sqrt{2}}}\partial _{t}\Psi _{1}\}=\Lambda
\Psi _{1}, }\eqno{[6a]}$$

$$
{{\sqrt{2}}\over{a\sqrt{sinh 2x}}}\{(\partial _{x}-i\partial
_{\theta })\Psi _{4}  \
-a[cos (\theta -ix)\partial _{y}+sin (\theta -ix)\partial _{z}]\Psi _{3}-%
{{a\sqrt{sinh 2x}}\over{\sqrt{2}}}\partial _{t}\Psi _{2}\}=\Lambda
\Psi _{2},\eqno{[6b]}$$
$$
{{\sqrt{2}}\over{a\sqrt{sinh 2x}}}\{(\partial _{x}-i\partial
_{\theta }+coth 2x)\Psi _{1}  \
-a[cos (\theta +ix)\partial _{y}+sin (\theta +ix)\partial _{z}]\Psi _{2}+%
{{a\sqrt{sinh 2x}}\over{\sqrt{2}}}\partial _{t}\Psi _{3}\}=\Lambda
\Psi _{3},\eqno{[6c]}$$
$$
{{\sqrt{2}}\over{a\sqrt{sinh 2x}}}\{(\partial _{x}+i\partial
_{\theta }+coth 2x)\Psi _{2}  \
+a[cos (\theta -ix)\partial _{y}+sin (\theta -ix)\partial _{z}]\Psi _{1}+%
{{a\sqrt{sinh 2x}}\over{\sqrt{2}}}\partial _{t}\Psi _{4}\}=\Lambda
\Psi _{4}.\eqno{[6d]}
 $$
  These are coupled equations involving three different components.
  The usual method to obtain solutions at this stage
is the separation of variables method.  Three of the variables,
$t,y,z$, define the Killing directions. This is exploited by
writing the solution as a product of exponentials in $y,z,t $
times a function of $ x$ and $\theta$.
$$ \Psi_{i} =e^{i(k_{t}t+k_{y}y+k_{z}z)}{\Psi_{i}} (x,\theta
).\eqno{[7]}$$ We take $ k_{y}=k cos (\phi), k_{z}=k sin(\phi)$,
since then we can absorb the variable $\phi$ in functions of the
remaining variables $x $ and $\theta$.

\noindent We note that these four equations are not similar in
form.
  The transformation $ \Psi_{1,2} = {{1}\over {
\sqrt{sinh{2x}}}} f_{1,2} $ is used for the upper components. This
transformation eliminates the $coth {2x}$ terms in the third and
the fourth equations.  With this transformation, the equations
read
$$
\{(\partial _{x}+i\partial _{\theta })\Psi _{3}  \
+iak[cos (\theta -\phi +ix)]\Psi _{4}-%
iak_t f_{1}\}=\Lambda {{a}\over{\sqrt{2}}}f_{1},\eqno{[8a]}$$

$$\{(\partial _{x}-i\partial
_{\theta })\Psi _{4}  \
-iak[cos (\theta -\phi-ix) ]\Psi _{3}-%
iak _{t}f_{2}\}=\Lambda  {{a}\over{\sqrt{2}}}f_{2},\eqno{[8b]}$$
$$
(-\partial _{x}+i\partial _{\theta })f_{1}  \ +iak[cos (\theta
-\phi+ix)]f_{2}+i {a k _t\Psi_{3}=  -\Lambda {{asinh
2x}\over{\sqrt{2}}}\Psi_{3}},\eqno{[8c]}$$
$$
(-\partial _{x}-i\partial _{\theta })f_{2}  \ -iak[cos (\theta
-\phi-ix)] f_{1}+i a k_{t}\Psi _{4}=-\Lambda {{asinh
2x}\over{\sqrt{2}}} \Psi _{4}. \eqno{[8d]}$$

\noindent
 We solve our equations in terms of $f_{1,2} $ and substitute these
expressions  in equations, given above. This substitution gives us
second order, but uncoupled equations for the lower components.
$$ \left( \partial _{xx}+\partial _{\theta \theta }+{{a^{2}}\over{2}}
[k^{2} \left(- cos [2(\theta -\phi )]-cosh 2x \right)
-(k_{t}^{2}+\Lambda^2) sinh 2x  ]\right)  \Psi _{3,4}=0 .\eqno{[9]}
$$ We can separate this equation into two ordinary differential
equations by the ansatz $\Psi _{3,4}= R(x) S(\theta-\phi)$.
Separation of equation(9) gives us two ordinary differential
equations.  The equation for $S$ reads
$$    \partial _{\Theta \Theta }S(\Theta )-\left({{a^{2}}\over {2}} k^{2}
 cos (2{\Theta  })
-n \right)  S(\Theta )=0 ,      \eqno{[10]}$$ where $(\theta -\phi
)=\Theta$. This equation is of the Mathieu type and the solution can
be written immediately.

$$ S(\theta) =C_1 Se(n, {{a^2k^2}\over {4}}, \theta-\phi)
+ C_2  So(n, {{a^2k^2}\over {4}}, \theta-\phi)\eqno{[11]}$$ The
solutions should be periodic in the angular variable $\Theta$. This
fact forces  $n$, the separation constant, to take discrete values.
It is known that the angular Mathieu functions satisfy an
orthogonality relation such that functions with different $n$ values
are perpendicular to each other. Here, we integrate the angular
variable from zero to $2\pi$. One can fix the normalization constant
according to the chosen normalization, whether it is according to
the McLachlan or the Morse-Stratton convention [32 ].

\noindent The equation for $R(x)$ reads
$$
\left\{ \partial _{xx}-[{{a^{2}\over{2}} }(k^{2}cosh
2x+(k_{t}^{2}+\Lambda^2)sinh 2x]+n\right\} R (x)=0 . \eqno{[12]}$$
whose solution can be reduced to the form
$$ R (x) = D_1 Se(n, A_6, i(x+b))+D_2 So(n, A_6, i(x+b)) . \eqno{[13]}$$
 Here $C_1,C_2,D_1,D_2$ are arbitrary constants.
The other constants will be defined below as we explain how one
can reduce our initial equation to give these solutions. The
solutions for the lower components $\psi_3, \psi_4 $ are given in
terms of sums over $n$ and integrals over $k, \phi,
 k_t$ we used in the separation ansatz. We think it is amusing to
 explain how these solutions are obtained. To get this solutions one
 has to go through several steps.

\noindent In reference books the solution of eq. (13) is not listed
as a Mathieu function. In fact, we find that the solution of this
equation is expressed in terms of double confluent Heun functions
[33 ].  We denote these functions as $H_D$ in our expressions.

$$
R(x) =C_1H_D\left(
0,{{a^{2}k^2}\over{2}}-n,a^{2}(\Lambda^{2}+k_{t}^{2}) ,{{
a^{2}k^{2}}\over{2}}+n,tanh x\right)  $$
$$+C_2H_D\left( 0,{{a^{2}k^2}\over{2}}-n,a^{2}(\Lambda^{2}+k_{t}^{2}) ,{{
a^{2}k^{2}}\over{2}}+n,tanh x\right) $$
$$\times \int {{{-dx}\over{H_D\left(  0,{{a^{2}k^2}\over{2}}-n,a^{2}(\Lambda^{2}+k_{t}^{2}) ,{{
a^{2}k^{2}}\over{2}}+n,tanh x\right)^2}}}  \eqno{[14]}$$ \noindent
Normally one takes the first function and discards the second
solution.

\noindent Reading through literature [34 ]  we suspect that the
solution we found can be expressed in terms of Mathieu functions
after performing proper transformations. We can show that we can
indeed express  the result in terms of Mathieu functions by using a
simpler independent variable than the one given in eq. (14) after
performing  few transformations. We define
$$
A_1={{-a^{2}(k_{t}^{2}+\Lambda^2)}\over{2}},\eqno{[15]}$$
$$A_2=-n \eqno{[16]}$$
$$
A_3=-{{a^{2}k^{2}}\over{2}},\eqno{[17]}$$ and use the transformation
$$
z=e^{-2x}.\eqno{[18]}$$ Then the differential operator in eq. (12)
is expressed as
$$
\left(4z^{2}\partial _{zz}+4z\partial
_{z}+({{A_3-A_1}\over{2}})z+A_5+ ({{A_3+A_1}\over{2}}) {{1}\over{z}}
\right) f=0 .\eqno{[19]}$$ This equation is still of the double
confluent Heun form, since it still has two irregular singularities
at zero and infinity. We define $ A_4= ({{A_3-A_1}\over{2}})$ and
$A_5= ({{A_3+A_1}\over{2}})$. If we take
$$
\sqrt{{{A_4}\over{A_5}}}u=z, \eqno{[20]}$$ and $$ w
={{1}\over{2}}(u+{{1}\over{u}}) \eqno{[21]}$$ and set $A_6=\sqrt{A_4
A_5}$ we get,
$$
\left((w^{2}-1)\partial _{ww}+w\partial
_{w}+({{A_6}\over{2}}w+{{A_2}\over{4}})\right) f =0. \eqno{[22]}$$
With the new transformations, we have traded the irregular
singularity at zero by two regular singularities at plus and minus
one.  This is the same singularity  structure of the Mathieu
equation. The solution of this equation is indeed expressible in
terms of Mathieu functions. It is given as:

$$
R(w) = Se\left(n,A_6,arccos
\sqrt{{{w+1}\over{2}}}\right)+So\left(n,A_6,arccos
\sqrt{{{w+1}\over{2}}}\right),\eqno{[23]}$$ where $A_6=
{{a^2}\over{4}} [k^4-(k_t^2+\Lambda^2)^2]^{1/2}. $ Going back
through the transformations we made, it is not hard to express
$arccos \sqrt{{{w+1}\over{2}}}$ in terms of our original variable
$x$ up to a constant, $i(x+b)$. We simply write $$ {{w+1}\over{2}}=
{{e^{-2x} \left(e^{2x} +
\sqrt{k^2-k_{t}^2-\Lambda^2}\over{(k^2+k_{t}^2+\Lambda^2}\right)^2}\over{
4 \sqrt{
{{k^2-k_{t}^2-\Lambda^2}\over{(k^2+k_{t}^2+\Lambda^2)}}}}}=cosh^2(x+b).\eqno{[24]}$$
Here
$$e^{-2b}=\sqrt{{{k^2-k_{t}^2-\Lambda^2}\over{k^2+k_{t}^2+\Lambda^2}}}.\eqno{[25]}$$
Taking the $arccos$ of this expression gives the result given in eq.
(13).

\noindent After these transformations,  we see that the solution
of the lower components of the Dirac equation can be expressed in
terms of functions that are regular at zero.  We can not say this
for the upper components, though. They are expressed in terms of
these solutions and their derivatives  divided by a function,
which blows up at zero. Even if we take the so called odd Mathieu
functions, which can be expanded in terms of hyperbolic sine
functions, their derivatives will be hyperbolic cosine functions.
To obtain the upper components we have to divide them by $\sqrt
{sinh {2x}}$ . Then these functions   will still blow at the
origin.

\noindent A finite scalar product can be defined around the origin
for these solutions in the form $$ \int \Psi_{i}^{*} \Psi_{i} \sqrt{
g} d\tau, i=1-4, \eqno{[26]}$$ in a finite domain, including the
origin. Here $d\tau$ is the volume element in our five dimensional
space. Repeated indices are not summed over. $\sqrt{ g}$ is the
square root of the determinant of the metric, necessary to get an
invariant volume element. The zero of the invariant  measure cancels
the singularity of the wave functions at the origin.

\noindent
 We find, however, that the solutions are not
normalizable as $x$ goes to infinity. Furthermore,  our metric has
curvature singularities at the  origin.  Although, using our new
measure, we can make all four components normalizable at the origin,
we still define our solutions for the domain $0<x<F$. We integrate
$x$ variable in the domain up to the point $x\leq F$, where $F$ is
the point the function starts to diverge.  Since our radial solution
is multiplied by the angular solution  and the exponential function
to make up the total solution of the Dirac equation, the
orthogonality of the angular solutions for different values of the
discrete $n$ makes our solutions orthogonal to each other. By
dividing by the appropriate factors, we can normalize them.

\noindent We note that  the domain where the solutions are
normalizable is restricted. These differential equations do not have
a meaning unless we define the boundary conditions the solutions
obey at the boundary of our domain. Since our system exists in an
odd dimensional manifold with a boundary, we have to study this
equation using spectral boundary conditions [9,10 ] .

\noindent The method used in applying these boundary conditions
requires first studying the little Dirac equation, the tangential
operator of the Dirac operator restricted to the boundary, where the
variable $x$ takes a fixed value $x_0$. We have to take $x_0$
greater than zero since the second fundamental form defined by our
choice of the normal to the boundary diverges at $x_0=0$ [18 ].

For this purpose we write the equations in the form given in
equations (8), i.e. after $\Psi_{1,2}$ are transformed to $f_{1,2}$.
$$
{{\sqrt{2}}\over{a}}\{i\partial _{\theta }\Psi _{3}  \ +iakcos
(\theta-\phi +ix_0)\Psi _{4}-
{{iak_t}\over{\sqrt{2}}}f_{1}\}=\lambda f _{1},\eqno{[27a]}$$
$$
{{\sqrt{2}}\over{a}}\{-i\partial _{\theta }\Psi _{4}  \ -iakcos
(\theta-\phi -ix_0)\Psi _{3} -{{iak_t}\over{\sqrt{2}}}f
_{2}\}=\lambda f_{2},\eqno{[27b]}$$
$$
{{\sqrt{2}}\over{a}}\{-i\partial _{\theta } f_{1}  \ -iak cos
(\theta -\phi+ix_0)f_{2} +{{iak_t
}\over{\sqrt{2}}}\Psi_{3}\}=\lambda \Psi_{3},\eqno{[27c]}$$
$$
{{\sqrt{2}}\over{a}}\{i\partial _{\theta }f _{2}  \ +iak cos
(\theta-\phi -ix_0)f _{1} +{{iak_t}\over{\sqrt{2}}}\Psi
_{4}\}=\lambda \Psi _{4}.
 \eqno{[27d]}$$
Here $\lambda$ is the eigenvalue of the little Dirac equation.

\noindent We could not obtain analytical solutions of these
equations in terms of known functions. We could not even write
uncoupled equations in the second order. One needs to go to fourth
order in derivatives to be able to write equations that involves a
single unknown function.

\noindent At this point we follow closely our references [6,7,8,31].
We formally expand our solutions at the boundary, fixed by two
values of $x_0$ in terms of eigenfunctions of the little Dirac
equations with both positive and negative eigenvalues $\lambda$. For
the lower components we have
$$\Psi_{3,4}^{\Lambda} (\Theta, x_0) = \sum_ {\lambda}h_{\lambda}(\Theta, x_0) \eqno{[28]}$$ for
fixed values of $k_{t}, k_{y}, k_{z} $.  We set
$$\Psi_{3,4}^{\Lambda} (\Theta, x_0)|_{\partial B}  = \sum_
{\lambda>0}h_{\lambda}(\Theta,x_0). \eqno{[29]}$$ The negative
$\lambda$ eigenvectors are all set to be zero at the boundary.

\noindent Then, we solve $f_{1,2}$ in terms of $\Psi_{3,4}$ using
the equations (8a, 8b) and fix the $x$ values to $x_0$  on the
boundary.  Note that since $x=x_0$, the derivative with respect to
$x$ is evaluated at this point as well as the terms without the $x$
derivative. We can, in general, use the expansion given for
$\Psi_{3,4}$ on the boundary, in terms of its eigenfunctions.
$$\Psi_{3,4}^{\Lambda} (\Theta, x_0) = \sum_
{\lambda}h_{\lambda,3,4}(\Theta,x_0) .\eqno{[30]}$$ This sum is over
all values of $\lambda$. In fixing the values of $f_{1,2} $ in terms
of $\Psi_{3,4}$ on the boundary, we use only part of the expansion
of  $\Psi_{3,4}^{\Lambda} (\Theta, x)$ where
$$\Psi_{3,4}^{\Lambda} (\Theta, x)|_{\partial B} = \sum_
{\lambda<0}h_{\lambda,3,4}(x_0,\Theta) .\eqno{[31]}$$  In other
words, we write $\Psi_{1,2}^{\Lambda} (\Theta, x)|_{\partial B}$
using the expressions obtained from $f_{1,2}$ in terms of the
negative $\lambda$ values of $\Psi_{3,4}^{\Lambda} (\Theta,
x)|_{\partial B}$. These boundary conditions are non local, but are
shown to be the only consistent ones for odd dimensional Euclidean
spaces by Atiyah-Patodi-Singer.

\vskip.3truein
 \noindent
 {\bf{Solutions in four dimensions}}
 \vskip.1truein
\noindent
 Here we repeat the calculations given above after
setting $k_t$ to zero.  Our aim is to show that the five dimensional
case, studied above, is not essentially different from the four
dimensional case. Since both solutions can be expressed in terms of
Mathieu functions we can compare the four and five dimensional
cases.  These equations with $\Lambda=0$ were studied in ref. [5 ].
The solutions obtained in this reference for the four dimensional
case can be summed to give the Green's function similar to the
calculation done in ref. [2 ]. This is not possible for the five
dimensional solution of ref. [5 ]. For the solutions with $\Lambda
\neq 0$ both in five and four dimensions this property does not
exist. The radial and the angular equations have different constants
which does not allow to obtain the Green's function of the equation
by using summation formulae of Mathieu functions [35].

\noindent Our solutions exist only in a finite domain for the
variable $x$, hence we have to use the appropriate boundary
conditions at this point. Using the orthogonality of the angular
Mathieu functions for different values of the discrete parameter
$n$, we can show the orthogonality of these solutions.

\noindent
 At this
point, although we can use local boundary conditions in this case,
we choose to use the spectral boundary conditions of
Atiyah-Patodi-Singer to conserve chirality and charge conjugation.
We are keen not to break chirality by hand, since in the standard
model this symmetry is expected to break spontaneously, resulting in
confinement.
 There are cases, however, to describe the correct physics, where
one may want to break this symmetry to force the system to one of
the broken phases.  Using chirality breaking boundary conditions may
be one way to achieve this task.  This is discussed in detail in
reference [36].  In this work, we do not want to break the chiral
symmetry of the model. We, therefore, adhere to the spectral
boundary conditions.

\noindent We write the system in the form $L \psi = \Lambda \psi $
  where  $\psi$ is a four component spinor, and try to obtain the
solutions for the different components. Then our equations are
similar to the ones given in the equations (6).  The only difference
is taking $k_{f}$ equal to zero. We see that three components are
still coupled in our equations.

\noindent The method of solution is exactly like the one used in the
previous section. To get our solutions we use the separation of
variables method. We write the solution as a product of exponentials
in $y,z $ times a function of $ x$ and $\theta$.
$$ \Psi_{i} =e^{i(k_{y}y+k_{z}z)}{\Psi_{i}} (x,\theta
) .\eqno{[32]}$$   The same transformations are used as those in the
five dimensional case to reduce $\Psi_{1,2}$ to $f_{1,2}$. We solve
our equations in terms of $f_{1,2} $ and substitute these
expressions in equations, given above.   We end up with second
order, but uncoupled equations for the lower components.
$$    \left( \partial _{xx}+\partial _{\theta \theta }-{{a^{2}}\over {2}}
[k^{2} \left( cos {2(\theta -\phi )}+cosh 2x \right) +\Lambda^2 sinh
2x ] \right)  \Psi _{3,4}=0 .      \eqno{[33]}$$ We can separate
this equation into two ordinary differential equations by the ansatz
$\Psi _{3,4}= R(x) S(\theta)$.  For $S(\Theta)$ we get
 an equation  of the Mathieu type and the solution can be written immediately.
$$ S(\theta) =C_1 Se[n, {{a^2k^2}\over {4}}, \theta-\phi] + C_2
So(n, {{a^2k^2}\over {4}}, \theta-\phi).\eqno{[34]}$$ \noindent
Solution for $R(x)$ can be reduced to
$$ R (x) = D_1 So(n, B, i(x+b'))+D_2Se(n, B, i(x+b')) ,\eqno{[35]}$$
   Here
$C_1,C_2,D_1,D_2$ are arbitrary constants.
$B={\sqrt{k^4-\Lambda^4}\over{2}}$. $b'$ is defined as in eq.(25)
with $k_t=0$ \noindent Just note that $n$ is the separation constant
which has to take discrete values to get a periodic solutions for
the angular Mathieu equation $S(\Theta)$. The solutions for the
lower components $\Psi_3, \Psi_4 $ are given in terms of sums over
$n$ and integrals over $k,\phi $ we used in the separation ansatz.

\noindent We find that we have the upper solutions,$\Psi_{1,2}$ are
divergent at the origin, whereas the lower ones are finite. Both of
our solutions diverge at infinity. For the same reasons as given for
the five dimensional case, we have to limit the domain of our
solutions at two finite values of x.

\noindent To impose these boundary conditions we need to write the
little Dirac equation, the Dirac equation restricted to the
boundary, where the variable $x$ takes a fixed value $x_0$. We
choose to write the equations in the form,
$$
{{\sqrt{2}}\over{a}}\{i\partial _{\theta }\Psi _{3}  \ +ikacos
(\theta-\phi +ix_0)\Psi _{4}\}=\lambda f_{1},\eqno{[36a]}$$

$$
{{\sqrt{2}}\over{a}}\{-i\partial _{\theta }\Psi _{4}  \ -iakcos
(\theta-\phi -ix_0)\Psi _{3}\}=\lambda f _{2},\eqno{[36b]}$$
$$
{{\sqrt{2}}\over{a}}\{(-i\partial _{\theta })f_{1}  \ -iak cos
(\theta -\phi+ix_0)f_{2} \}=\lambda \Psi_{3},\eqno{[36c]}$$
$$
{{\sqrt{2}}\over{a}}\{(i\partial _{\theta })f _{2}  \ +iakcos
(\theta-\phi -ix_0) f_{1} \}=\lambda \Psi _{4}.\eqno{[36d]}
$$
Here $\lambda$ is the eigenvalue of the little Dirac equation.

\noindent We could not obtain analytical solutions of these
equations in terms of known functions. This is the same result as in
the five dimensional case. For us it seems very curious being able
to solve similar system of partial differential equations, but not
even being able to decouple them when this system reduces to
ordinary differential equations on the boundary. One possible
explanation is that $\theta-\phi \pm ix$ act as $z$ and
  $\bar z$ of complex variables.  Sometimes it is easier to find functions of this pair as solutions
   is easier than a function of a single real variable.
Actually for the full Dirac equation, Sucu and \"Unal [3] find
solutions in a closed form. Same technique, however, does not seem
to work when $x=x_0$.

\noindent  From  equations (36) we see that we can take $\Psi_{4} $
as the complex conjugate of $\Psi_{3} $ and $ f_{2} $ as the complex
conjugate of $ f_{1} $. We take $ f_{1}= p_1+iq_1 $ and $\Psi_{3}=
p_3+iq_3 $. We end up with coupled differential equations for these
functions.One notes, however, that separating the real and imaginary
parts of each solution, we get
 $$
\partial _{\Theta \Theta }p_{1} +ak[-sin{\Theta} cosh {x_0}-(sin^2 {\Theta} sinh^2{x_0}
+cos^2 {\Theta} cosh^2{x_0}) ] p_{1} -ak cos{\Theta} sinh{x_0} q_{1}
= {\lambda^2} p_{1},     \eqno{[37]}$$

$$
\partial _{\Theta \Theta }q_{1} +ak[sin{\Theta} cosh {x_0}-(sin^2 {\Theta} sinh^2{x_0}
+cos^2 {\Theta} cosh^2{x_0}) ] q_{1} -ak cos{\Theta} sinh{x_0} p_{1}
= {\lambda^2} q_{1}.   \eqno{[38]}$$

\noindent We see that the eigenvalue of the little Dirac equation
comes only  quadratically, showing a symmetry for its positive and
negative values. This will make the $\eta$ invariant, defined as
$$\eta = lim_{s\rightarrow 0}\sum _{\lambda_i} sign(\lambda_{i}) |\lambda_{i}|^{-s}, \eqno{[39]}$$
which is needed for an index calculation, zero.   We also see a
symmetry between $p_1$ and $q_1$, namely $p_1(\Theta) = q_1
(-\Theta) .$ If we, instead, eliminate $p_{1}, q_{1}$ and write our
equations for $p_{3}, q_{3}$, we get exactly the same equations.

\noindent We expand our solutions at the boundary, fixed by two
values of $x_0$ in terms of eigenfunctions of the little Dirac
equations with both positive and negative eigenvalues $\lambda$.
$$\Psi_{i}^{\Lambda} (\Theta, x_0) = \sum_ {\lambda}g_{i,\lambda}(\Theta, x_0) \eqno{[40]}$$ for fixed values of $ k_{y}, k_{z} $.
We set
$$\Psi_{3,4}^{\Lambda} (\Theta, x )|_{\partial B}  = \sum_ {\lambda>0}g_{\lambda,3,4}(\Theta, x_0). \eqno{[41]}$$
The negative $\lambda$ eigenvectors are all set to be zero at the
boundary.

The boundary conditions on the upper components are imposed exactly
in the same manner as explained in the five dimensional case, namely
we solve for $f_{1,2}$ in terms of $\Psi_{3,4}$ using the equations
$$
{{\sqrt{2}}\over{a}}\{(\partial _{x}+i\partial _{\theta })\Psi _{3}
\ +a[cos (\theta -\phi+ix)]\Psi _{4}\}=\Lambda f _{1},\eqno{[42]}$$

$$
{{\sqrt{2}}\over{a}}\{(\partial _{x}-i\partial _{\theta })\Psi _{4}
\ -a[cos (\theta - \phi -ix)]\Psi _{3}\}=\Lambda f
_{2},\eqno{[43]}$$ and fix the $x$ values to $x_0$  on the boundary.

\noindent
 We can in general use the expansion given for $\Psi_{3,4}$ on the boundary,
in terms of its eigenfunctions, eqn. (40). This sum is over all
values of $\lambda$. In fixing the values of $f_{1,2} $ in terms of
$\Psi_{3,4}$ on the boundary, we use only the part where
$\lambda<0$.
 These
boundary conditions are non local, but they  respect self
adjointness   and conserve $\gamma^5$ and charge conjugation
symmetry.

\vskip.2truein
 \noindent
 {\bf{Conclusion}}
 \vskip.1truein
\noindent
 Here we tried to give solutions of the Dirac equation in
five dimensions for the Nutku helicoid metric in a bounded region.
We found out that they can be reduced to Mathieu functions, which is
also the case in four dimensions. We imposed  formally [8,31] the
non-local spectral boundary conditions of Atiyah-Patodi and Singer
[9,10,6,7] on these solutions, which are the only  correct boundary
conditions in odd dimensions.  Our main goal in this paper is to
define the solutions of the Dirac equation in the background of a
incomplete metric, namely the Nutku helicoid solution consistently.

\noindent A related work would be to calculate the index of the
Dirac operator in four dimensions in this background. From our
solutions we can  calculate both the bulk and the surface term
easily.  From the form of  the little Dirac equation, we see that
the $\eta$ invariant is zero. We could not obtain analytical
solutions of the little Dirac equation, though. We give examples of
the numerical solutions we found of the zero eigenvalue equations of
the little Dirac equation in Figure 1. These pictures do not
correspond to any of the functions we encounter in the literature
[32,33,35,37]. Therefore, we know that the zero mode solutions of
the little Dirac equation exist. We, however, do not know their
analytical expression, hence their number. As a result, we could not
calculate the index in this paper.\vskip.2truein

 \noindent
 {\bf{Acknowledgement}}: We thank Profs. John Roe,
Yavuz Nutku, Ay\c se Bilge and Ne\c se \" Ozdemir for
correspondence, for discussions and scientific assistance throughout
this work. The work of M.H. is also supported by TUBA, the Academy
of Sciences of Turkey. This work is also supported by TUBITAK, the
Scientific and Technological Council of Turkey.

\vskip.5truein

 \noindent {\bf{References}} \vskip.2truein
1. Nutku Y 1996 {\it{Phys.Rev. Lett.}} {\bf {77}} 4702

2. Aliev A N, Horta\c csu M, {Kalayc\i} J and Nutku Y 1999
{\it{Class. Quantum Grav.}} {\bf{16}} 631

3. Sucu Y and \" Unal N 2004 {\it{Class. Quant. Grav.}} {\bf{21}}
1443

4. Villalba V M 2005 {\it{J.Phys.: Conf. Ser.}} {\bf{24}} 136

5. Birkandan T and Horta\c csu M  2007 {\it{J.Phys. A }} {\bf{40}}
1105, e-Print Archive: gr-qc/0607108 and corrigendum to be published

6. Horta\c csu M, Rothe K D, Schroer B 1980 {\it{Nucl. Phys.}}
{\bf{B171}} 530

7. Horta\c csu M 1983 {\it{Lettere al Nuovo Cim.}} {\bf{36}} 109

8. Abrikosov, jr. A A, Wipf A 2007 {\it{J. Phys. A: Math. Theor.}}
{\bf{40}}  5163

9. Atiyah M F, Patodi V K and Singer I M 1975 {\it{Math. Proc.
Camb.Phil. Soc.}} {\bf{77}} 43

10. Atiyah M F, Patodi V K and Singer I M 1975 {\it{Math. Proc.
Camb.Phil. Soc.}} {\bf{77}} 405

11. Peeters K and Waldron A 1999 {\it{JHEP}} 9902:024; e-Print
Archive: hep-th/9901016

12. Roe J 1988 {\it{J. Differential Geom.}} {\bf{27}} 87

13. Roe J 1988 {\it{J. Differential Geom.}} {\bf{27}} 115

14. Falomir H 1997 e-Print Archive: physics/9705013

15. Atiyah M F and Bott R 1964, {\it{Index Theorem for Manifolds
with Boundary in : Differential Analysis }}(Bombay Colloquium)
Oxford Univ. Press

16. Gilkey P B,  Kirsten K and J.H. Park J. H.  2005
 {\it{J.Phys.A}} {\bf{38}}8103,
e-Print Archive: math-ph/0406028

17. Gilkey P B 1984 {\it{Invariance Theory, the Heat Equation and
the Atiyah-Singer Index Theorem}}  Publish or Perish Inc., Delaware,
USA.

18. Eguchi T, Gilkey P.B and Hanson A 1980, {\it{Physics Reports}}
{\bf{C66}} 213

19. Gilkey P B 1984 {\it{Invariance Theory, the Heat Equation and
the Atiyah-Singer Index Theorem}}  Publish or Perish Inc., Delaware,
USA. p.248 Lemma 4.1.6

20. Esposito G {\it{Dirac Operators and Spectral Geometry}}
Cambridge Lect.Notes Phys.12:1-209,1998 Cambridge Univ. Press,
Cambridge, England

21. Popescu A S  2007, "Dimension Embedded in Unified Symmetry",
e-Print Archive: gr-qc/0704.2670

22. Gibbons G W  and Rychenkova P  2000 {\it{ J. Geometry and
Phys.}} {\bf{32}} 311

23. Valent G and Ben Yahia H 2007 {\it{Class. and Quant. Grav.}}
{\bf{24}} 255

24. Valent G 2004 {\it{Commun. Math. Phys.}} {\bf{244}} 571

25. Valent G 2005 {\it{Int. J. Mod. Phys. A}} {\bf{20}} 2500

26. Newman E T and Penrose R 1962 {\it{J. Math. Phys.}} {\bf{3}} 566

27. Newman E T and Penrose R 1962 {\it{J. Math. Phys.}} {\bf{4}} 998

28. Goldblatt E 1994 {\it{Gen.Rel. Grav.}} {\bf{26}} 979

29. Goldblatt E 1994 {\it{J. Math. Phys.}} {\bf{35}} 3029

30. Aliev A N and Nutku Y 1999 {\it{Class. Quantum Grav.}} {\bf{16}}
1892

31. Abrikosov jr. A A 2006 {\it{J.Phys.A}} {\bf{39}} 6109 e-Print
Archive: hep-th/0512311

32. Guti\'{e}rrez Vega J C  2003 "Theory and Numerical Analysis of
Mathieu Functions", Tecnol\'{o}gico de Monterrey, M\'{e}xico report
and the references given in this paper.

33. Ronveaux A (ed) 1995 {\it{ Heun's Differential Equations}}
(Oxford: Oxford University Press)

34. Schmidt D and Wolf G  in:  Ronveaux A (ed) 1995 {\it{ Heun's
Differential Equations}}  (Oxford: Oxford University Press)

35.  Morse P M and Feshbach H 1953 Methods of Theoretical Physics
(McGraw Hill) p.1421

36. Wipf A and Duerr S 1995 {\it{Nucl. Phys.B}} {\bf{443}} 202

37. Slavyanov S  Yu, in :Ronveaux A (ed) 1995 {\it{ Heun's
Differential Equations}} (Oxford: Oxford University Press). p.95

\vfill\eject

{\bf{ FIGURE CAPTIONS}}

\vskip .2truein Figure 1 : Numerical solution for zero mode
solutions of $p_1$ and $q_1$ in four dimensions for $x_0=0.005, a=1,
k=1$.

\vfill\eject
   \midinsert
   \vskip .75truein
   \centerline{\BoxedEPSF{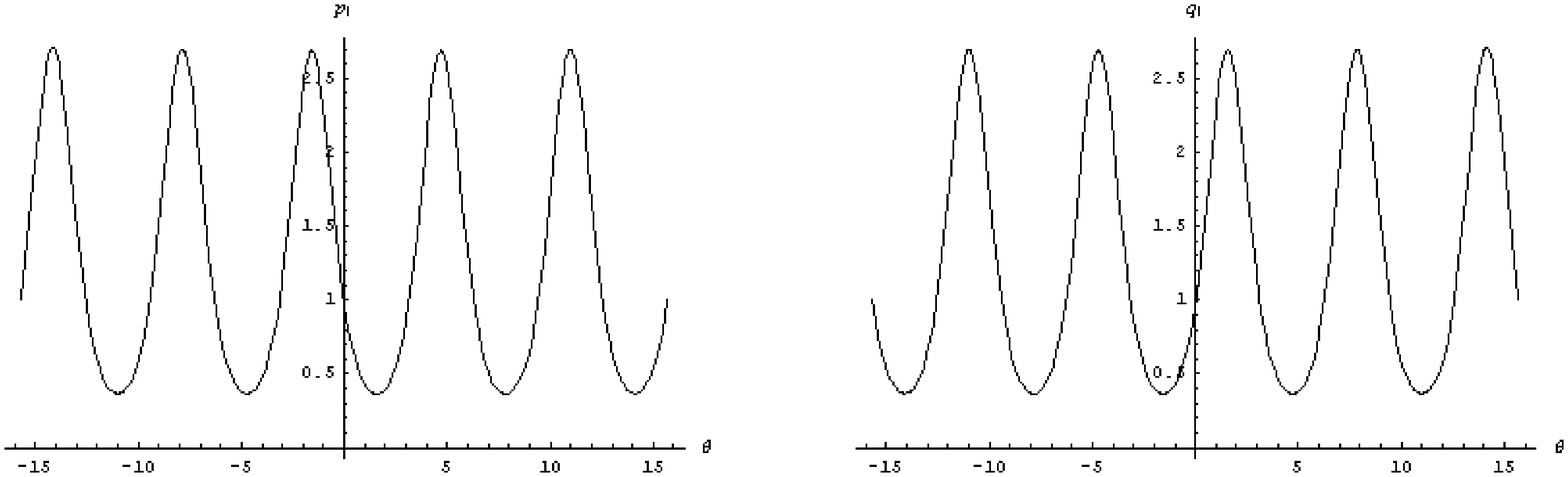}}
   \vskip .2truein
   \centerline {\bf { Figure 1}}
   \vskip .5truein
   \endinsert

\vfill\eject
\end